\documentclass[11pt]{article}
\usepackage{moriond,epsfig}

\def\Journal#1#2#3#4{{#1} {\bf #2}, #3 (#4)}

\def\PRL{\em Phys. Rev. Lett.}
\def\PRD{{\em Phys. Rev.} D}

\def\be{\begin{equation}}
\def\ee{\end{equation}}
\def\bea{\begin{eqnarray}}
\def\eea{\end{eqnarray}}

\begin{document}
\vspace*{4cm}
\title{PROBING NON-GAUSSIANITIES ON LARGE SCALES IN WMAP5 AND 
         WMAP7 DATA USING SURROGATES}

\author{C. R\"ATH$^1$, G. ROSSMANITH$^1$, G. E. MORFILL$^1$, A. J. BANDAY$^{2,3}$, K. M. G\'{O}RSKI$^{4,5}$}
\address{$^1$Max-Planck-Institut f\"ur extraterrestrische Physik, Giessenbachstr. 1, 85748 Garching, Germany\\
 $^2$Centre d'Etude Spatiale des Rayonnements, 9, Av du Colonel Roche, 31028 Toulouse, France\\
 $^3$Max-Planck-Institut f\"ur Astrophysik, Karl-Schwarzschild-Str. 1, 85741 Garching, Germany\\
 $^4$Jet Propulsion Laboratory, California Institute of Technology, Pasadena, CA 91109, USA\\
  $^5$Warsaw University Observatory, Aleje Ujazdowskie 4, 00 - 478 Warszawa, Poland
}

\maketitle\abstracts{Probing Gaussianity represents one of the key questions in modern cosmology, 
because it allows to discriminate between different models of inflation. We test for large-scale
non-Gaussianities in the cosmic microwave background (CMB) in a model-independent way. 
To this end, so-called first and second order surrogates are generated by first shuffling the Fourier phases 
belonging to the scales not of interest and then shuffling the remaining phases for the length scales under study. 
Using scaling indices as test statistics we find highly significant signatures for both non-Gaussianities and asymmetries 
on large scales for the WMAP data of the CMB. We find remarkably similar results when analyzing different ILC-maps based
on the WMAP five and seven year data.
Such features being independent from the map-making procedure 
would disfavor the fundamental principle of isotropy as well as canonical single-field slow-roll inflation -  
unless there is some undiscovered systematic error in the collection or reduction of the 
CMB data or yet unknown foreground contributions.}

\section{Introduction}

One of the key questions in cosmology is to probe the Gaussianity of the primordial density fluctuations, 
because it allows to discriminate between different models of inflation. 
While the simplest model of inflation, namely single-field slow-roll inflation\cite{Guth81}, 
predicts that the temperature fluctuations of the cosmic microwave background (CMB) 
correspond to a (nearly) Gaussian, homogeneous, and isotropic 
random field, more complex models may give rise to non-Gaussianity (NG). Models in which the Lagrangian is a general 
function of the inflaton and powers of its first derivative can lead to scale-dependent non-Gaussianities, 
if the sound speed varies during inflation\cite{Armendariz99,Garriga99}. Similarly, string theory models that give rise to large non-Gaussianity 
have a natural scale dependence \cite{Loverde08a}.
Possible deviations from Gaussianity  have been investigated in 
studies based on e.g. the WMAP data of the CMB 
(see e.g. Komatsu et al.\cite{Komatsu09a} 
and references therein) and claims for the detection of 
non-Gaussianities and other anomalies, like hemispherical power asymmetry, the 'axis of evil' ,
the Cold Spot etc.  have been made.
However, most of the tests on non-Gaussianities do not take into account any possible 
scale-dependency. 
In this contribution we apply the formalism of surrogate data sets, which was recently  adapted to CMB data analysis \cite{Raeth09},
to test for non-Gaussianities on large scales in the WMAP data.
Special emphasis is put on a comparison of the WMAP five year data with the WMAP seven year data.

\section{Data Sets}

We use the foreground-cleaned  
Internal Linear Combination (ILC) maps generated and provided  by
the WMAP-team on the basis of the five year (WMAP5)\cite{Gold09a}
and seven year (WMAP7)\cite{Gold10a} data.
For comparison we also analysed the five year ILC-map (NILC5) produced by 
Delabrouille et al.\cite{Delabrouille09},
which was generated pursuing a needlet-based approach for removing the 
foreground contributions. 

\section{Surrogates and Scaling Indices}

The model-independent test for scale-dependent non-Gaussianities 
is based on the use of so-called surrogate data sets. As test statistics 
for NGs we use scaling indices. Both methods were introduced
and described in detail previously \cite{Raeth09,Raeth07,Rossmanith09}.
Here, we briefly review the main points of the two formalisms.

\subsection{Surrogates}\label{subsec:surro}
\begin{figure}
\begin{center}
\psfig{figure=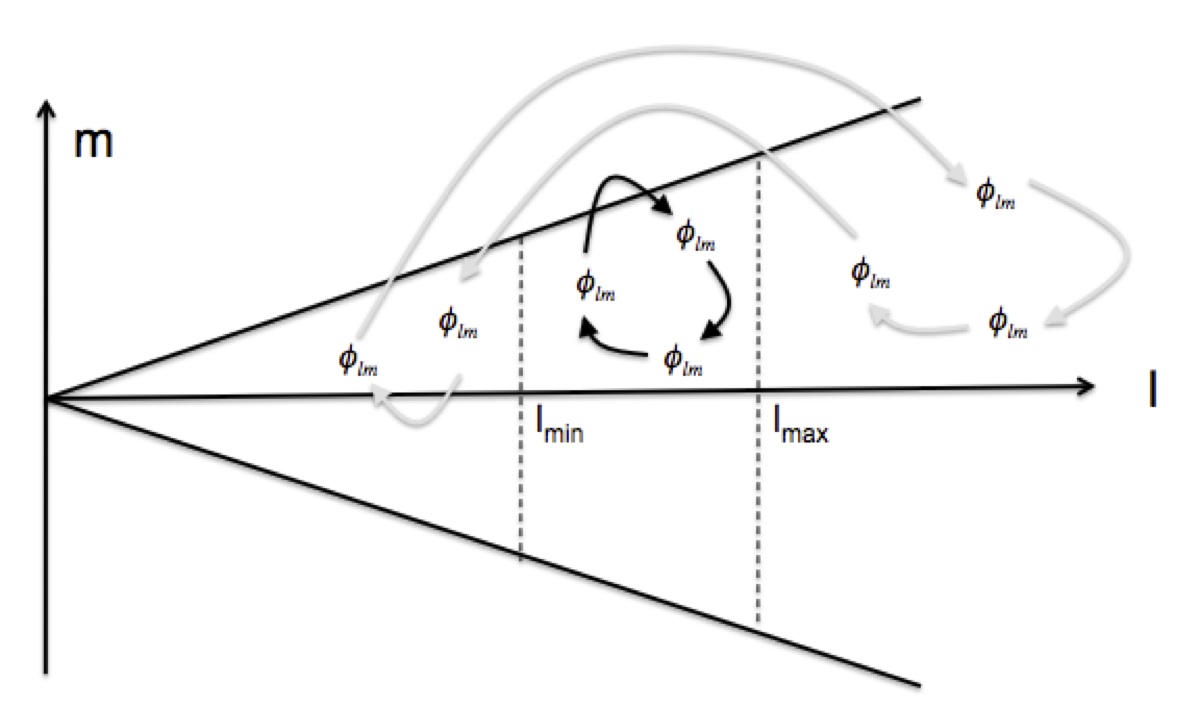,height=5.0cm}
\end{center}
\caption{Schematic view of the shuffling procedure on the $l$-$m$-plane for generating the first and second order surrogates. The gray arrows indicate
the phase permutations for obtaining first order surrogates and the black arrows show the shuffling for making the second 
order surrogates. 
\label{fig:shuffling}}
\end{figure}

Consider a CMB map $T(\theta,\phi)$,
where $T(\theta,\phi)$ is Gaussian distributed, which 
can easily be achieved by a rank-ordered remapping
of the temperatures onto a Gaussian distribution.
We calculate the Fourier 
transform of $T(\theta,\phi)$. The complex valued coefficients $a_{lm}$,
$a_{lm} = \int d\Omega_n T(n) Y^{*}_{lm}(n)$
can be written as
$a_{lm} = | a_{lm} | e^{i \phi_{lm}} $ 
with  $\phi_{lm}=\arctan \left( Im(a_{lm}) / Re(a_{lm} )  \right)$.
The linear or Gaussian properties of the underlying random field 
are contained in the absolute values $ | a_{lm} | $, 
whereas all higher order correlations (HOCs) -- if present -- 
are encoded in the phases $\phi_{lm}$ and the correlations among them. 
To ensure that the distribution of the phases is 
uniform, we also perform a rank ordered remapping of the phases $\phi_{lm}$.
To test for scale-dependent NGs we first generate a first order surrogate map, 
in which any phase correlations 
for the scales, which are not of interest (here: the small scales), 
are randomized.  This is achieved by a random shuffle of the phases  $\phi_{lm}$ 
for $l > l_{cut}, 0 < m \le l$, where  $l_{cut}=20$ throughout this study and
by performing an inverse Fourier transformation.
Second, $N$ (here: $N=500$) 
realizations of second order surrogate maps 
are generated for the first order surrogate map, in which the remaining 
phases $\phi_{lm}$  with $1 < l \le l_{cut}, 0 < m \le l$ are shuffled while the 
already randomized phases for the small scales are preserved. In Fig. \ref{fig:shuffling}
the phase shuffling procedure for generating first and second order surrogates is 
schematically visualized.

\subsection{Scaling Indices}\label{subsec:sim}
To compare the two classes of surrogates, we calculate local statistics
in the spatial domain, namely scaling indices (SIM)  
as described previously\cite{Raeth07,Rossmanith09}.
In brief, scaling indices estimate
local scaling properties of a point set $P$.
The spherical CMB data can be represented as a
three-dimensional point distribution $P=\vec{p_i} = (x_i,y_i,z_i), i=1,\ldots,N_{pixels}$ 
by transforming the temperature fluctuations into a radial jitter. 
For each point $\vec{p_i}$ the local weighted cumulative point distribution
$\rho$ is calculated   $\rho(\vec{p_i},r) = \sum_{j=1}^{N_{pixels}} e^{-(\frac{d_{ij}}{r})^2}$
with $d_{ij} = \| \vec{p_i} - \vec{p_j} \| $.
The weighted scaling indices $\alpha(\vec{p_i},r)$ are then obtained by calculating
the logarithmic derivative of $\rho(\vec{p_i},r)$ with respect to $r$,
$\alpha(\vec{p_i},r) = \frac{\partial \log \rho(\vec{p_i},r)}{\partial \log r}$. 
Using the above-given expression for the  local weighted cumulative point distribution
$\rho$, the following analytical formula for the scaling index $\alpha$
\begin{equation}
  \alpha(\vec{p_i},r) = \frac{\sum_{j=1}^{N} 2 (\frac{d_{ij}}{r})^2
                                             e^{-(\frac{d_{ij}}{r})^2}}
                             {\sum_{j=1}^{N} e^{-(\frac{d_{ij}}{r})^2}}
\end{equation}
is obtained. For each pixel we calculated scaling indices for ten different 
scales, $r_1=0.025$,\ldots,$r_{10}=0.25$ in the notation of R\"ath et al.\cite{Raeth07}.

\section{Results}

\begin{figure}
\begin{center}
\psfig{figure=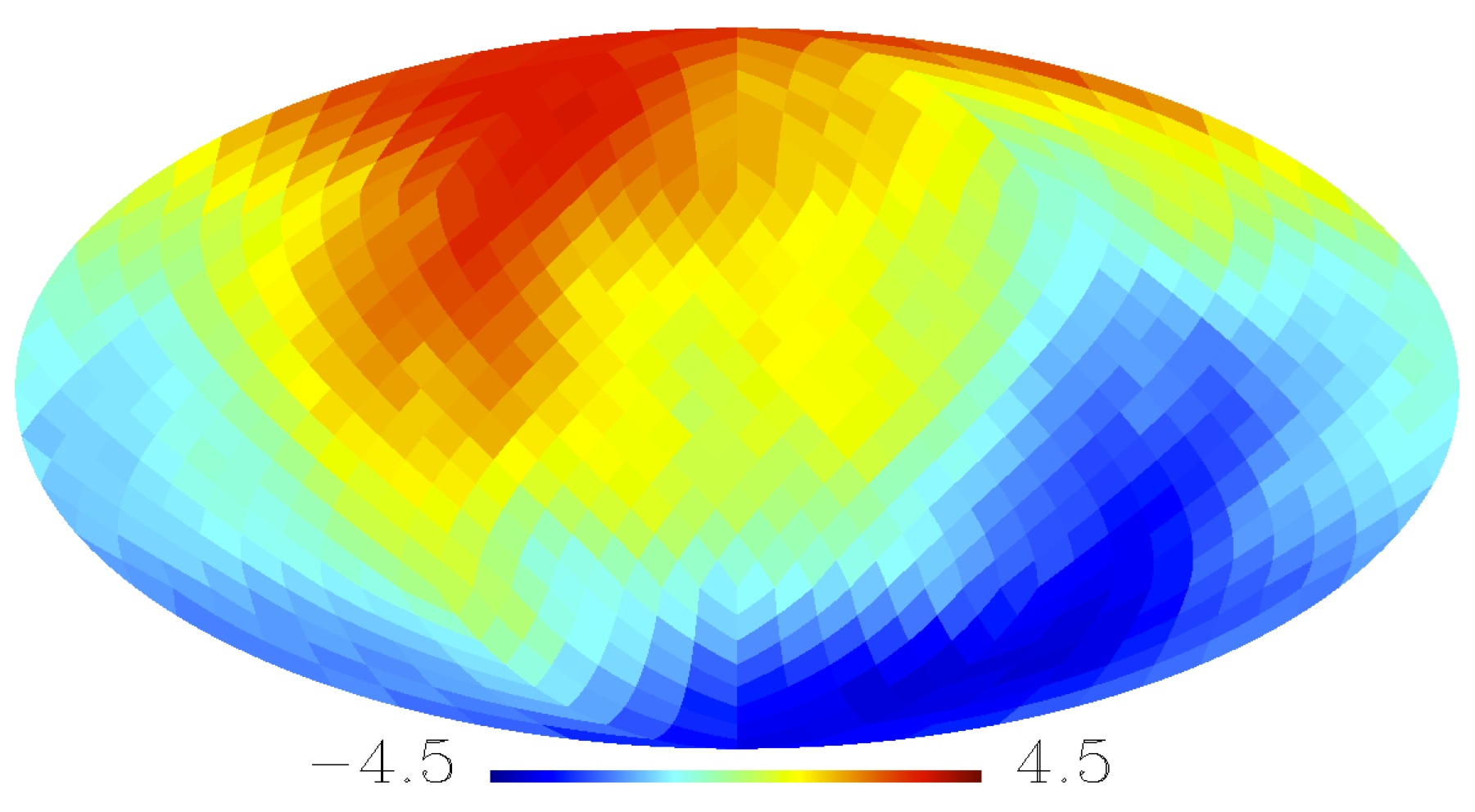,height=2.85cm}
\psfig{figure=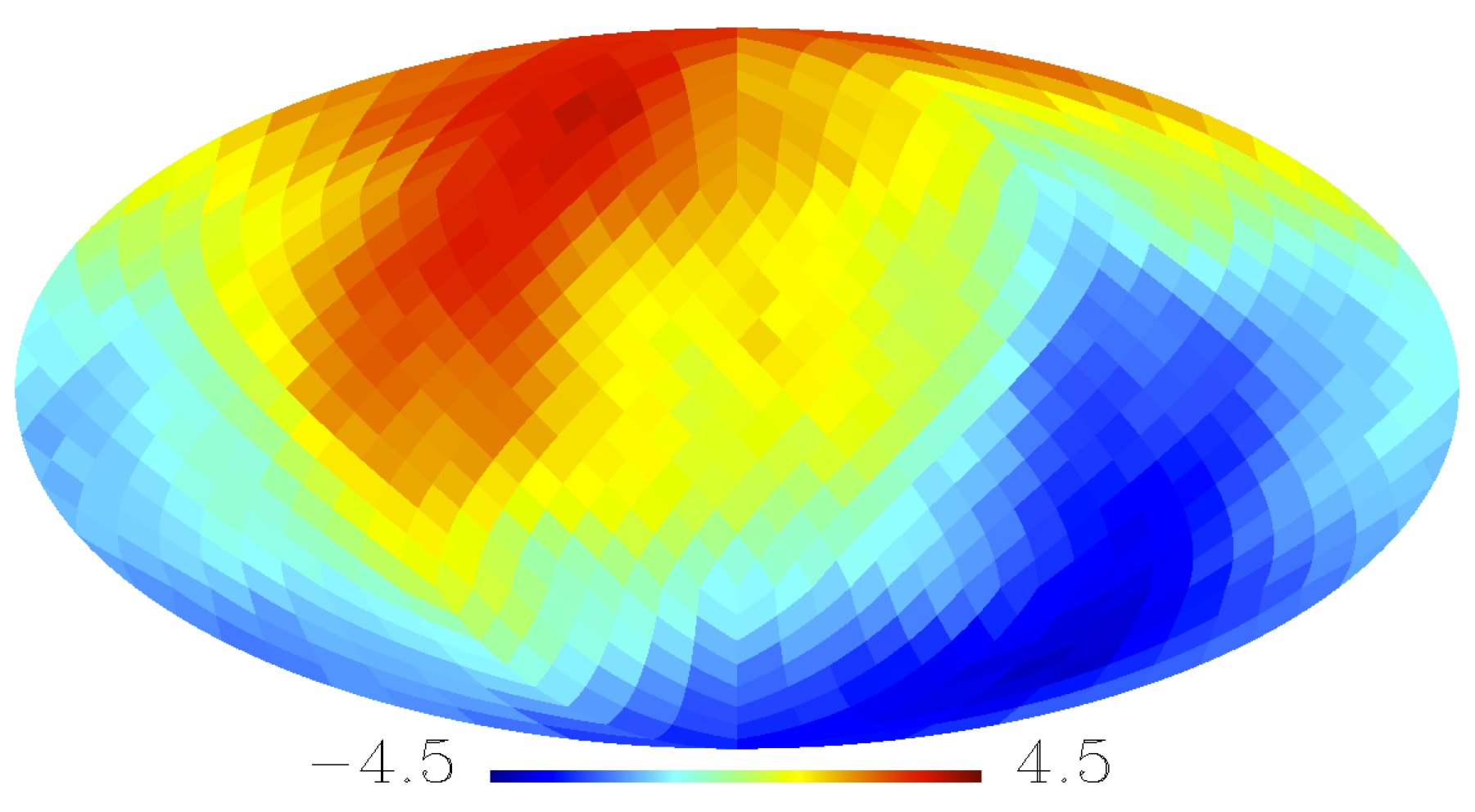,height=2.85cm}
\psfig{figure=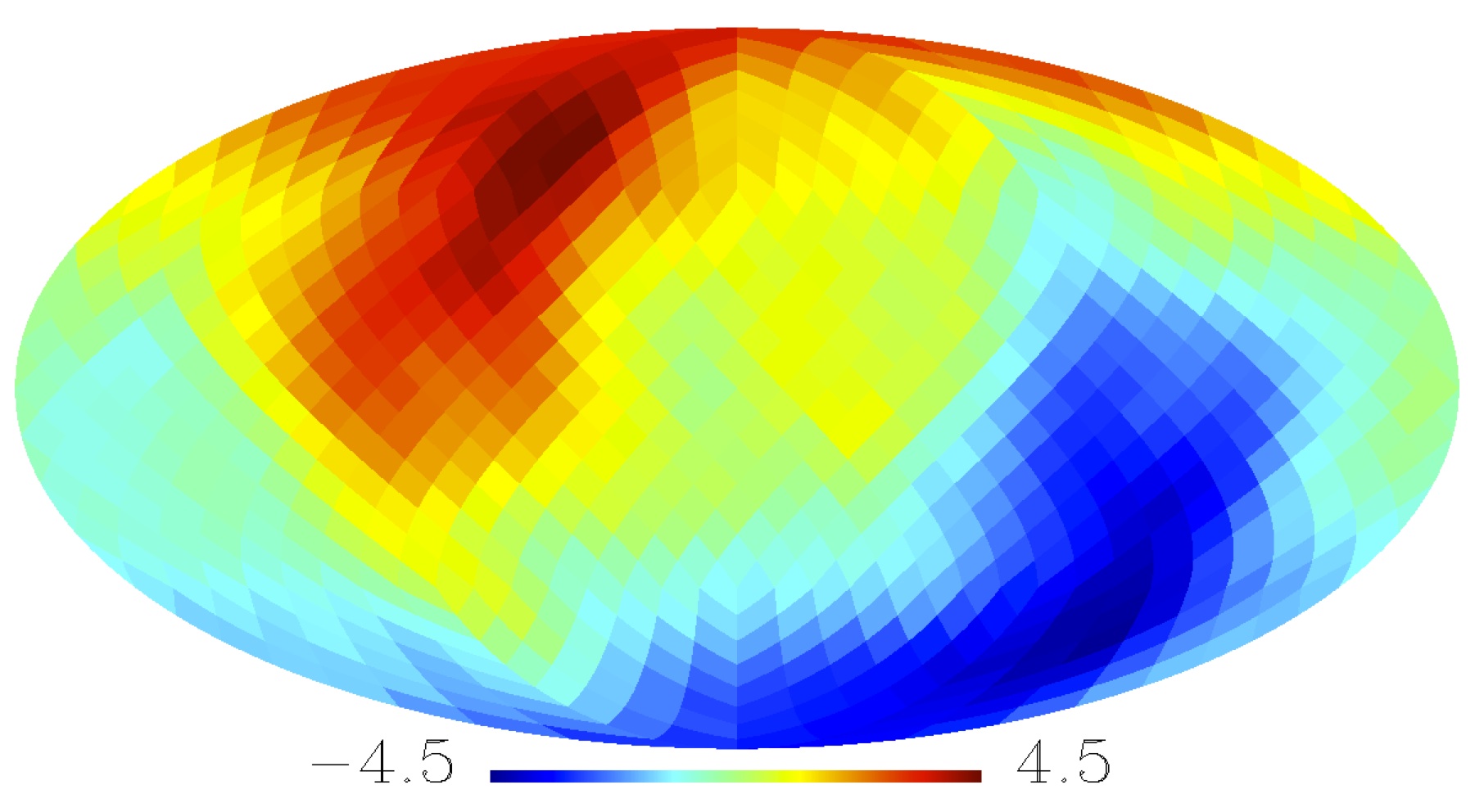,height=2.85cm}
\end{center}
\caption{Deviation $S$  as derived from rotated upper hemispheres for $\langle\alpha(r_{10})\rangle$ for 
the WMAP5 ILC map (left), the  WMAP7 ILC map (middle) and the needlet-based ILC map (right).
\label{fig:sig_maps}}
\end{figure}

For each scale we calculate the mean ($\langle \alpha \rangle$) 
of the scaling indices  $\alpha(\vec{p_i},r)$ derived from a set of pixels belonging 
to rotated hemispheres. 
The differences of the two classes of surrogates 
are then quantified by the $\sigma$-normalised deviation 
$S(Y)= (Y_{surro1} - \langle Y_{surro2} \rangle)/ \sigma_{Y_{surro2}}$,
$Y=\langle \alpha (\vec{p_i},r_j) \rangle$, $j=1,\ldots,10$, 
surro1: first order surrogate, surro2: second order surrogate.
Fig. \ref{fig:sig_maps} 
shows the deviation $S(\langle \alpha(r_{10}) \rangle)$ as derived from 
pixels belonging to  the respective upper hemispheres for $768$ rotated reference 
frames for the three ILC-maps under study. The z-axis  of the respective rotated reference 
frame pierces the center of the respective colour-coded pixel.

\begin{figure}
\begin{center}
\psfig{figure=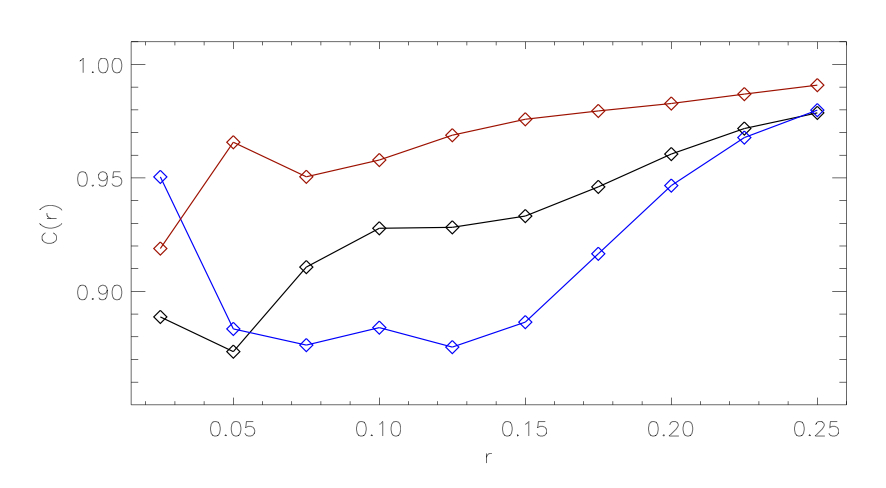,height=4.0cm}
\psfig{figure=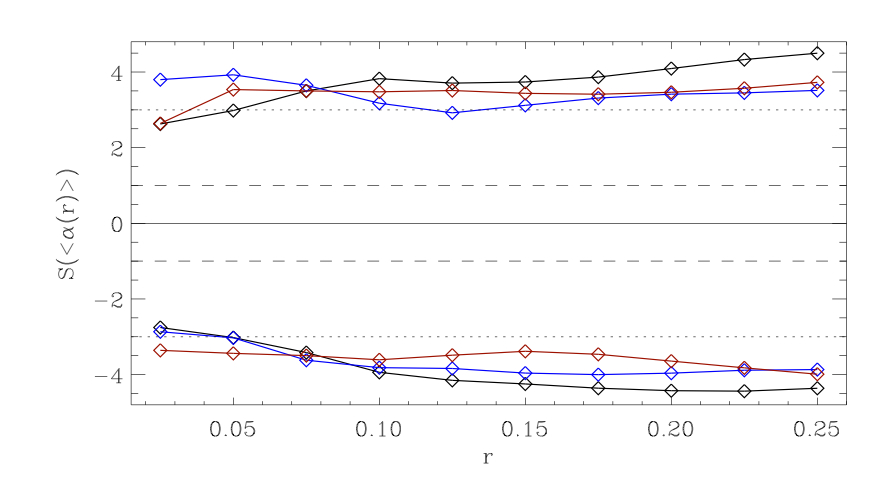,height=4.0cm}
\end{center}
\caption{Left side: Cross correlation coefficient $C(r)$ for the $S(Y)$-maps as a function of the scaling range $r$. 
Black: NILC vs. WMAP5, 
blue: NILC vs. WMAP7 and red: WMAP5 vs. WMAP7. Right side: Minimal and maximal 
values of the $\sigma$-normalized deviations $S(Y)$ for the rotated hemispheres for each scaling range $r$. 
Black: NILC, blue: WMAP5 and red: WMAP7.
\label{fig:cross_corr}}
\end{figure}

Statistically significant signatures
for non-Gaussianity and ecliptic hemispherical asymmetries 
become immediately obvious, 
whereby the patterns of asymmetry remain remarkably similar 
for the three maps. 
Interestingly enough, we obtain slightly larger deviations for the WMAP7 map 
($S_{min}=-3.99$, $S_{max}=3.73$) as compared to the WMAP5 map
($S_{min}=-3.87$, $S_{max}=3.51$). We find the largest deviation  
($S_{min}=-4.36$, $S_{max}=4.5$) for the 
NILC map, which can be considered as a more  precise full-sky CMB temperature map
than the ILC maps generated by the WMAP team\cite{Delabrouille09}.
Thus, the level of non-Gaussianity systematically increases, when the underlying CMB-map becomes 
less noisy and less foreground contaminated.\\ 
To quantify the similarity of the patterns of asymmetry in the three maps  we calculate the cross-correlation
$C(r)$ of the $S(Y)$-maps derived from the three input maps as a function of the scaling range $r$. 
The results are shown in fig.  \ref{fig:cross_corr} (left). For each scaling range $r$ the three 
maps are highly correlated among each other with $C(r)$ always being larger than $0.87$. The 
highest correlations are found for the largest scales $r$, where $C(r)$ reaches values of  $0.98$ 
and more. Thus, on all scales $r$ the patterns of asymmetry are very similar for each input map.
On the right hand side of fig. \ref{fig:cross_corr} we show the minimum and maximum of $S$ as a 
function of $r$. Except for the smallest $r$'s we obtain for each map 
stable $3 \sigma$-deviations for both extrema, where -- once again -- the NILC map always yields the largest
deviations for scaling ranges $r>0.1$. 
 
\section{Conclusions}

In conclusion, we detect highly significant signatures for asymmetries and non-Gaussianities
for large scales ($l<20$) in the WMAP five and seven year data.  
The increase of the signal with decreasing noise as well as the very high correlations 
between the significance maps points towards an intrinisic nature of the detected anomalies, 
which are independent of the map making procedure.
Such features would disfavor the fundamental principle of isotropy as well as canonical single-field slow-roll inflation -  
unless there is some undiscovered systematic error in the collection or reduction of the 
CMB data or yet unknown foreground contributions.
Thus, further tests are required to rule out other systematic effects as origin of the detected anomalies.

\section*{Acknowledgments}
Many of the results in this contribution have been obtained using 
HEALPix\cite{Gorski05}. 
We acknowledge the use of LAMBDA. Support for LAMBDA is provided by the 
NASA Office of Space Science. Finally, CR would like to thank the 
organisers for a very enjoyable meeting.

\end{document}